\documentclass[doublecol]{epl2}

\usepackage{bm}
\usepackage{graphicx}
\usepackage{amsmath}
\usepackage{amssymb}

\title{Anomalous Hall effect with massive Dirac fermions}

\author{I. A. Ado\inst{1} \and I. A. Dmitriev\inst{2,3} \and P. M. Ostrovsky\inst{2,4} \and M. Titov\inst{1}}
\shortauthor{I. A. Ado \etal}

\institute{
 \inst{1} Radboud University, Institute for Molecules and Materials - NL6525 AJ Nijmegen, The Netherlands\\
 \inst{2} Max Planck Institute for Solid State Research - Heisenbergstr.\,1, 70569 Stuttgart, Germany\\
 \inst{3} A.\,F.\,Ioffe Physico-Technical Institute - 194021 St.\,Petersburg, Russia\\
 \inst{4} L.\,D.\,Landau Institute for Theoretical Physics RAS - 119334 Moscow, Russia
}

\pacs{72.10.-d}{Theory of electronic transport; scattering mechanisms}
\pacs{72.25.-b}{Spin polarized transport}
\pacs{72.10.Bg}{General formulation of transport theory}

\abstract{
Anomalous Hall effect arises in systems with both spin-orbit coupling and magnetization. Generally, there are three mechanisms contributing to anomalous Hall
conductivity: intrinsic, side jump, and skew scattering. The standard diagrammatic approach to the anomalous Hall effect is limited to computation of ladder
diagrams. We demonstrate that this approach is insufficient. An important additional contribution comes from diagrams with a single pair of intersecting
disorder lines. This contribution constitutes an inherent part of skew scattering on pairs of closely located defects and essentially modifies previously
obtained results for anomalous Hall conductivity. We argue that this statement is general and applies to all models of anomalous Hall effect. We illustrate it
by an explicit calculation for two-dimensional massive Dirac fermions with weak disorder. In this case, inclusion of the diagrams with crossed impurity lines
reverses the sign of the skew scattering term and strongly suppresses the total Hall conductivity at high electron concentrations.
}

\begin{document}

\maketitle

Many ferromagnetic materials exhibit a finite Hall effect, i.e.\ transverse voltage in response to a current, without applying external magnetic field. This
phenomenon is commonly referred to as the anomalous Hall effect (AHE) \cite{Nagaosa2010rev}. Two important ingredients of AHE are magnetization and spin-orbit
interaction. The former breaks time-reversal symmetry and exerts a force acting on electron spins while the latter couples the spins to orbital degrees of
freedom thus giving rise to the transport effect.

AHE can also occur as a result of valley or isospin polarization rather than ordinary ferromagnetism \cite{Perel1971}. The spin-orbit coupling can also be of a
more general form as it is, e.g.\ in graphene \cite{Novoselov2005, Geim2007} where the role of spin is played by the sublattice index. An important part of the
anomalous Hall signal originates in the Berry curvature thus being of a topological origin \cite{Haldane1988}. It is, therefore, natural that the discovery of
materials like graphene and topological insulators \cite{Hasan2010rev, Qi2011rev}, which are characterized by non-trivial Berry phase of quasiparticles, has
considerably broadened the interest to AHE from both theory and experiment \cite{Inoue2006, Niu2007, Liu2008, Ezawa2012, Zhang2013, Mak2014, Pan2015}.

Despite the long history \cite{Karplus1954, Kondo1962, Nozieres1973} and high experimental relevance of AHE, its theoretical description is a challenging task
often leading to confusions. In modern literature, two common approaches based on the Boltzmann kinetic equation and Kubo-St\v{r}eda diagrammatic formalism are
discussed. Boltzmann equation provides an intuitive quasiclassical approach to the effect \cite{Sinitsyn2008rev, Nagaosa2010rev} but requires an accurate
account of several mechanisms of Hall conductivity: intrinsic, side-jump, and skew-scattering. Intrinsic AHE is attributed to topological properties of the band
\cite{Jungwirth02} and is thus independent of disorder. Skew scattering is due to the asymmetry in the impurity scattering cross-section and side jump refers to
the transverse displacement of an electron being scattered. An alternative microscopic Kubo-St\v{r}eda formalism is more systematic but less intuitive
\cite{Sinitsyn2007, Nunner07}. In this Letter we will employ the latter approach but also discuss the results in terms of the three quasiclassical mechanisms.

The conductivity of a good metal at low temperature, such that direct contributions of inelastic scattering processes can be disregarded, can be expanded in
powers of the large metallic parameter $p_0 l$ (here $p_0$ is the Fermi momentum and $l$ is the electron transport mean free path). Within the standard model of
weak Gaussian disorder, the dissipative component of the conductivity tensor behaves as  $\sigma_{xx} \propto p_0 l$ in agreement with the Drude theory, while
the anomalous Hall conductivity, $\sigma_{xy}\propto (p_0 l)^0$, remains constant in the clean limit $p_0 l \to \infty$. Beyond the Gaussian approximation, an
additional scattering parameter $l^* \lesssim l$ (related to the third moment of random potential distribution) arises leading to an enhanced skew scattering,
$\sigma_{xy} \propto p_0 l^*$ \cite{Kooi1954, Nagaosa2010rev,Sinitsyn2008rev}. In the limit of rare strong impurities, this contribution can develop into the
resonant skew scattering \cite{Fert1972, Coleman1985, Fert2011, Ferreira2014}.

\begin{figure}
\onefigure{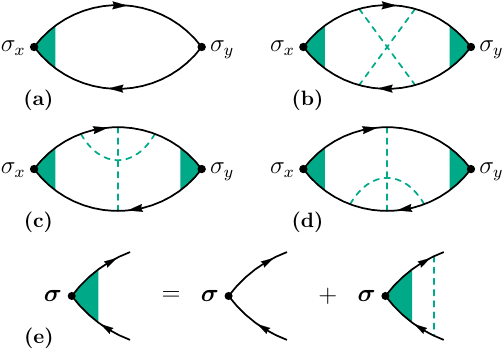}
\caption{Diagrams for the anomalous Hall conductivity. Standard non-crossing approximation (a) yields Eq.\ (\protect\ref{DD}). The X diagram (b) is of the same
(zeroth) order in disorder strength as all the non-crossing diagrams and corrects the result leading to Eq.\ (\protect\ref{DX}). The $\Psi$ diagram (c-d) has
potentially the same order of magnitude but cancels for the massive Dirac model. Vertex correction involves the sum of ladder diagrams (e), see Eq.\
(\protect\ref{dressed}).}
\label{fig:bubbles}
\end{figure}

In this Letter we consider AHE in the model of weak Gaussian disorder using the diagrammatic Kubo-St\v{r}eda formalism \cite{Streda}. It is well known that in a
good metal, $p_0 l \gg \hbar$, diagrams with intersecting impurity lines generically yield an extra smallness of the order $(p_0 l)^{-1}$. That is why earlier
calculations of anomalous Hall conductivity \cite{Sinitsyn2006, Sinitsyn2007, Nunner07} employed the non-crossing approximation including only the ladder
diagram Fig.~\ref{fig:bubbles}a. However, we show that, contrary to the common knowledge, the ``X and $\Psi$ diagrams'' with two intersecting impurities,
Fig.~\ref{fig:bubbles}(b-d), are also crucially important for AHE. Despite impurity-line crossing, they contribute to the result in the leading order $(p_0 l)^0$.

%%%%%%%%%%%%%%%%%%%%%%%%%%%%
%%%% fig:bubbles
%%%%%%%%%%%%%%%%%%%%%%%%%%%%
%%%%%%%%%%%%%%%%%%%%%%%%%%%%%

We argue that importance of these diagrams is a general feature of AHE and related phenomena. Indeed, the Hall conductivity in the non-crossing approximation
contains, in particular, the skew scattering contribution \cite{Sinitsyn2007}. Since any individual weak impurity has a symmetric scattering profile in the Born
approximation, the skew scattering occurs only due to pairs of impurities located at a distance less or of the order of the Fermi wave length, $p_0 |\mathbf{r}
- \mathbf{r}'|\lesssim \hbar$. Such closely positioned scatterers cannot be treated quasiclassically as isolated objects. Therefore a correct treatment of
scattering on the double defects should necessarily include the X and $\Psi$ diagrams. In the framework of the Boltzmann kinetic equation, inclusion of these
crossed diagrams corresponds to a proper calculation of the full two-impurity scattering amplitude in the collision integral.

To illustrate the general statement formulated above, we resort to the minimal model for AHE: 2D massive Dirac fermions in the presence of scalar Gaussian
disorder,
\begin{subequations}
\label{model}
\begin{gather}
 H = v\,\bm{\sigma} \mathbf{p} + m \sigma_z + V(\mathbf{r}), \label{Ham} \\
 \langle V(\mathbf{r}) V(\mathbf{r}') \rangle
  = 2\pi\alpha (\hbar v)^2 \delta(\mathbf{r} - \mathbf{r}'), \label{VV}
 \qquad
 \langle V \rangle
  =0.
\end{gather}
\end{subequations}
Here $\bm{\sigma} = (\sigma_x, \sigma_y)$ is the vector of Pauli matrices, $\mathbf{p}$ is the momentum operator, $v$ is the characteristic velocity, $\alpha$
is a dimensionless parameter characterizing disorder strength, and the brackets $\langle \dots \rangle$ stand for the averaging over disorder realizations. 

Aside from random potential, the Hamiltonian (\ref{Ham}) contains exactly the two required terms: spin-orbit coupling and magnetization. It is thus a generic
model of AHE \cite{Sinitsyn2006}. Apart from that, such Hamiltonian occurs on the surface of a 3D topological insulator subject to a constant magnetization $m$
(e.g., due to proximity to a ferromagnet). Another possible realization is graphene aligned on the surface of BN (in this case $m$ may have large-scale spatial
variations). The same Hamiltonian describes electron properties of a narrow HgTe quantum well (linearized Bernevig-Hugh-Zhang model \cite{BHZ}). In the last two
cases, two mutually time-reversed valleys described by Eq.~(\ref{Ham}) arise, hence $\sigma_{xy}$ has to be treated as the valley-Hall response. For HgTe, it
also corresponds to spin-Hall since the valleys have opposite magnetic polarization.

We will evaluate $\sigma_{xy}$ using Kubo-St\v{r}eda formula \cite{Streda}. The Hall conductivity is expressed as $\sigma_{xy} = \sigma_{xy}^\text{I} +
\sigma_{xy}^\text{II}$ with
\begin{equation}
 \sigma_{xy}^\text{I}
  = \frac{e^2}{h} \mathop{\mathrm{Tr}} \left< \sigma_x G^R \sigma_y G^A \right>,
 \qquad
 \sigma_{xy}^\text{II}
  = e c\, \frac{\partial n}{\partial B}.
\end{equation}
Here $G^{R,A}$ are the retarded and advanced Green functions at the Fermi energy, $n$ is the total electron concentration, and $B$ is the external magnetic
field. The first term describes contribution of conduction electrons near the Fermi surface while $\sigma_{xy}^\text{II}$ accounts for the contribution of the
entire Fermi sea and has a topological origin \cite{Jungwirth02}.

When the Fermi energy $\epsilon$ is inside the spectral gap, $|\epsilon| < |m|$, Hall conductivity is dominated by the second term $\sigma_{xy}^\text{II} =
-(e^2/2h) \mathop{\mathrm{sign}} m$ \cite{Haldane1988, Ludwig1994, Ostrovsky2007}. As the energy is increased above $m$, the conductivity $\sigma_{xy}$ decays
from $\pm 1/2$ to zero.  We will analyze this dependence as a function of the energy and mass (assuming from now on $\epsilon > m > 0$) in the leading (zeroth)
order in the disorder strength $\alpha$. The topological contribution $\sigma_{xy}^\textrm{II}$ is negligible outside the gap, $\sigma_{xy}^\textrm{II} \propto
\alpha$, hence our task is reduced to the analysis of the contribution $\sigma_{xy}^\textrm{I}$.

The standard approach to computing the leading contribution at the Fermi surface is the non-crossing approximation. It amounts to summing up the ladder diagrams
depicted in Fig.\ \ref{fig:bubbles}(a,e). Solid lines in the diagrams correspond to the disorder-averaged Green function and the dashed lines are disorder
correlators (\ref{VV}). The logic behind this formalism is based on the fact that the average Green function has a sharp maximum at the Fermi momentum $p_0 =
v^{-1} \sqrt{\epsilon^2 - m^2}$. The contribution of a diagram is maximized when all Green functions are taken at $p = p_0$. Each crossing of impurity lines
reduces the phase space available for the corresponding diagram and thus makes it smaller by the factor $(p_0 l)^{-1} \propto \alpha$.

\begin{figure}
\onefigure{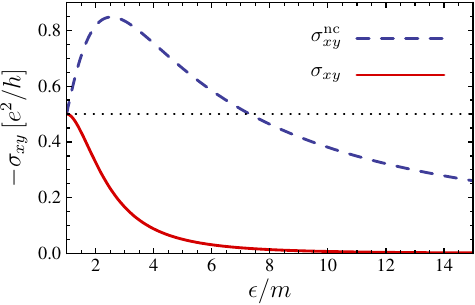}
\caption{Full anomalous Hall conductivity $\sigma_{xy}$, Eq.\ (\ref{DX}), including the contribution of the X and $\Psi$ diagrams (solid line). Anomalous
Hall conductivity $\sigma^\text{nc}_{xy}$, Eq.\ (\ref{DD}), calculated within the standard non-crossing approximation (dashed line).
}
\label{fig:result}
\end{figure}

\begin{figure}
\onefigure{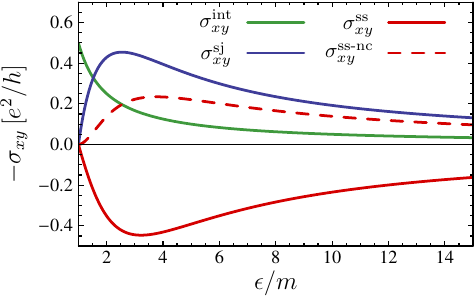}
\caption{
Intrinsic (\protect\ref{int}), side-jump (\protect\ref{sj}), and skew-scattering (\protect\ref{ss}) components of the anomalous Hall conductivity (solid lines).
Dashed line shows the skew scattering part in the non-crossing approximation (\protect\ref{ss-nc}).}
\label{fig:intsjss}
\end{figure}

Calculation of $\sigma_{xy}$ for the model (\ref{model}) in the non-crossing approximation was performed in Refs.\ \cite{Sinitsyn2006, Sinitsyn2007} with the
following result:
\begin{equation}
\label{DD}
 \sigma^\text{nc}_{xy}  = -\frac{4e^2}{h}\, \frac{\epsilon m (\epsilon^2 + m^2)}{(\epsilon^2 + 3 m^2)^2},
 \qquad
 \epsilon > m.
\end{equation}
This result is plotted in Fig.~\ref{fig:result} with the dashed line. In the kinetic equation formalism, this Hall conductivity is separated into
intrinsic, side-jump, and skew-scattering parts [see Eqs.\ (\ref{intsjss}) and Fig.\ \ref{fig:intsjss} below]. In the limit $\epsilon \gg m$ the value of
$\sigma^\text{nc}_{xy}$ decays as $m/\epsilon$.

%%%%%%%%%%%%%%%%%%%%%%%%%%%%
%%%% fig:result
%%%%%%%%%%%%%%%%%%%%%%%%%%%%
%%%%%%%%%%%%%%%%%%%%%%%%%%%%%

However, as we demonstrate below, the non-crossing approximation is incomplete. The contribution of X and $\Psi$ diagrams, Fig.~\ref{fig:bubbles}(b-d), is of
the same order $\alpha^0$ and modifies the result leading to
\begin{equation}
\label{DX}
 \sigma_{xy}
  = -\frac{8 e^2}{h}\, \frac{\epsilon m^3}{(\epsilon^2 + 3 m^2)^2},
 \qquad
 \epsilon > m,
\end{equation}
which is manifestly different from Eq.\ (\ref{DD}) as illustrated in Fig.\ \ref{fig:result}. In particular, the Hall conductivity $\sigma_{xy}$ decays as
$(m/\epsilon)^3$ in the limit $\epsilon \gg m$ in sharp contrast to the behavior of $\sigma^\text{nc}_{xy}$. The contribution of X and $\Psi$ diagrams is
important in the whole parameter range $\epsilon > m$. We also show that, for a given model, there exist no other contributions to $\sigma_{xy}$ in the leading
order in $\alpha$.

\begin{figure}
\onefigure{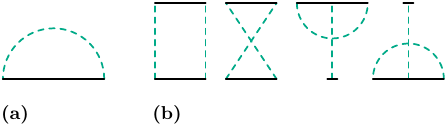}
\caption{Electron self energy (a) provides logarithmic renormalization of $\epsilon$ and $m$, while the disorder strength $\alpha$ is renormalized due
to vertex diagrams (b), see Eq.\ (\protect\ref{RG}). Imaginary part of the self energy (a) is finite and gives rise to the electron scattering rate in the Born
approximation. It is explicitly included in the average Green function (\protect\ref{Gep}).}
\label{fig:RG}
\end{figure}

Let us now turn to the details of the calculation. We start with the computation of the disorder-averaged Green function. For weak disorder, $\alpha
\ll 1$, the lowest order self energy (Born approximation) is given by the diagram depicted in Fig.\ \ref{fig:RG}a. The real part of this diagram diverges logarithmically
with the band width $\Lambda_0$. This divergence indicates that the Born approximation may be insufficient for the massive Dirac model (\ref{model}). Instead,
we perform an accurate summation of leading logarithmic contributions with the help of the renormalization group (RG) procedure \cite{Ludwig1994, Aleiner06,
Ostrovsky06, Evers08}. Apart from the self energy diagram Fig.\ \ref{fig:RG}a, RG also includes logarithmically divergent vertex diagrams Fig.\ \ref{fig:RG}b.
As a result of renormalization, the parameters of the model depend on the value of the running ultraviolet cutoff $\Lambda$,
\begin{equation}
 -\frac{d \alpha}{d \ln \Lambda}
  = 2 \alpha^2,
 \quad
 -\frac{d \ln\epsilon}{d \ln \Lambda}
  = \alpha,
 \quad
 -\frac{d \ln m}{d \ln \Lambda}
  = -\alpha.
\label{RG}
\end{equation}

%%%%%%%%%%%%%%%%%%%%%%%%%%%%
%%%% fig:RG
%%%%%%%%%%%%%%%%%%%%%%%%%%%%
%%%%%%%%%%%%%%%%%%%%%%%%%%%%%

Starting from the band width $\Lambda_0$ and bare parameters of the model, which are temporarily denoted as $\epsilon_0$, $m_0$, and $\alpha_0$, we solve the one-loop RG equations (\ref{RG}) and obtain the observable (renormalized) values
\begin{gather}
 \alpha
  = Z \alpha_0,
 \qquad
 \epsilon
  = Z^{1/2} \epsilon_0,
 \qquad
 m
  = Z^{-1/2} m_0, \\
 Z(\Lambda)
  = [1 - 2 \alpha_0 \ln(\Lambda_0/\Lambda)]^{-1}. \label{Z}
\end{gather}
The RG process has to stop when the running scale reaches $\Lambda = \sqrt{\epsilon^2 - m^2}$. At this point all logarithmic contributions are taken into account. We
assume that the disorder strength $\alpha$ is still small after the renormalization.

From now on we use only renormalized parameters of the model and hence disregard the real part of the Born self energy Fig.\ \ref{fig:RG}a. The imaginary part
of the self energy results in small imaginary contributions to energy and mass yielding the retarded Green function
\begin{equation}
\label{Gep}
 G^R_\mathbf{p}
  = \frac{\epsilon (1 + i \pi \alpha/2) + m \sigma_z (1 - i \pi \alpha/2)  + \bm{\sigma} \mathbf{p}}
         {\epsilon^2 - m^2 - p^2 + i \pi \alpha(\epsilon^2 + m^2)}.
\end{equation}
Advanced Green function is the hermitian conjugate of the above expression. Here and below we let $\hbar = v = 1$ to simplify notations.

We use the averaged Green function (\ref{Gep}) to calculate a single-impurity vertex correction,
\begin{gather}
 2 \pi \alpha \int \frac{d^2 p}{(2\pi)^2} G^A_\mathbf{p} \sigma_x G^R_\mathbf{p}
  = A \sigma_x + B \sigma_y, \label{dress} \\
 A  = \frac{\epsilon^2 - m^2}{2(\epsilon^2 + m^2)},
 \qquad
 B  = -\frac{\pi \alpha \epsilon m}{\epsilon^2 + m^2}.
\end{gather}
Note that the diagonal term $A$ is of the zeroth order in the parameter $\alpha$ while the factor $B \propto \alpha$ is small. Summing up the vertex ladder, see
Fig.\ \ref{fig:bubbles}e, we include the $B$ term only once. This yields the following expression for the dressed current operator:
\begin{equation}
\label{dressed}
 \bar\sigma_x
  = F \sigma_x + F^2 B \sigma_y,
 \qquad
 F
  = \frac{1}{1 - A}
  = \frac{2(\epsilon^2 + m^2)}{\epsilon^2 + 3 m^2}.
\end{equation}
The non-crossing approximation to $\sigma_{xy}$ is the result of adding one last ladder rung (\ref{dress}) to the already dressed current vertex $\bar\sigma_x$
and taking the trace with the bare current operator $\sigma_y$. This results in $\sigma_{xy}^\text{nc} = F^2 B/\pi \alpha$, which coincides with
Eq.~(\ref{DD}) obtained previously in Ref.\ \cite{Sinitsyn2006}.

To understand why the result of Eq.\ (\ref{DD}) is incomplete, one has to figure out which characteristic momenta provide the main contribution to it. This has
been already investigated in detail in Refs.\ \cite{Sinitsyn2007, Sinitsyn2008rev}, where the same problem had been addressed in the eigenbasis of the bare
Hamiltonian. This basis contains two eigenstates $|\pm\rangle$ per each momentum corresponding to the two bands of the Dirac Hamiltonian $\pm\sqrt{p^2 + m^2}$. Velocity operator does not commute with the Hamiltonian and hence has finite off-diagonal matrix elements in this basis. Setting all the Green functions on the mass
shell $\epsilon = \sqrt{p^2 + m^2}$ (i.e., completely disregarding the negative ``positron'' band) correctly reproduces $\sigma_{xx}$ but leads to vanishing
$\sigma_{xy}$.

In order to obtain a finite Hall conductivity in the zeroth order in $\alpha$, one has to allow for exactly one internal momentum in the diagram laying outside
the mass shell. In the bare bubble with no impurities, this is achieved by taking one Green function, which connects the two current vertices, away from the
mass shell (i.e., two off-diagonal matrix elements of velocity are involved). This yields the \textit{intrinsic} contribution to AHE. Similarly, the
\textit{side-jump} term corresponds to diagrams containing only one off-diagonal matrix element of velocity. In these diagrams, the off-mass-shell Green
function connects the off-diagonal current vertex to an impurity. Finally, the \textit{skew-scattering} diagrams contain two diagonal current vertices while the
off-mass-shell Green function connects two impurities. Within the non-crossing approximation, these three components of the anomalous Hall conductivity are
given by \cite{Sinitsyn2006, Sinitsyn2007}
\begin{subequations}
\label{intsjss}
\begin{align}
 \sigma_{xy}^\text{int}
  &= -\frac{e^2}{h}\, \frac{m}{2\epsilon}, \label{int} \\
 \sigma_{xy}^\text{sj}
  &= -\frac{e^2}{h}\, \frac{2m (\epsilon^2 - m^2)}{\epsilon (\epsilon^2 + 3 m^2)}, \label{sj} \\
 \sigma_{xy}^\text{ss-nc}
  &= -\frac{e^2}{h}\, \frac{3m (\epsilon^2 - m^2)^2}{2\epsilon (\epsilon^2 + 3 m^2)^2}. \label{ss-nc}
\end{align}
\end{subequations}
They are plotted in Fig.\ \ref{fig:intsjss}. The Green function taken away from the mass shell decays rapidly in real space. Therefore the latter
skew-scattering contribution is due to close pairs of impurities (separated by a distance $d\lesssim p_0^{-1}$ smaller or of the order of the Fermi wave
length). This means that crossing the corresponding impurity lines does not introduce an extra smallness! That is why the X and $\Psi$ diagrams depicted in
Fig.~\ref{fig:bubbles}(b-d) are of the same (zeroth) order in $\alpha$ as all the non-crossing terms and provide an important additional part of skew
scattering.

%%%%%%%%%%%%%%%%%%%%%%%%%%%%
%%%% fig:intsjss
%%%%%%%%%%%%%%%%%%%%%%%%%%%%
%%%%%%%%%%%%%%%%%%%%%%%%%%%%%

In order to calculate the contribution of X and $\Psi$ diagrams we resort to the real space representation. The diagrams of Fig.~\ref{fig:bubbles}(b-d) are
given by the following integrals:
\begin{subequations}
\label{XPsi}
\begin{align}
 \sigma_{xy}^\text{X}
  &= \frac{e^2}{h} \left(2\pi\alpha\right)^2 \int d^2r \mathop{\mathrm{Tr}} \Bigl( J^x_\mathbf{r} G^R_{-\mathbf{r}} J^y_\mathbf{r} G^A_{-\mathbf{r}} \Bigr), \\
 \sigma_{xy}^\Psi
  &= \frac{e^2}{h} \left(2\pi\alpha\right)^2 \int d^2r \mathop{\mathrm{Tr}} \Bigl(
      J^x_\mathbf{r} G^R_{-\mathbf{r}} G^R_\mathbf{r} J^y_{-\mathbf{r}} + \text{h.c.}
      %+ J^x_\mathbf{r} J^y_{-\mathbf{r}} G^A_{\mathbf{r}} G^A_{-\mathbf{r}}
    \Bigr),
\end{align}
\end{subequations}
where the vector $\mathbf{r}$ connects the impurities. We define $\mathbf{J}_\mathbf{r}$ as the Fourier transform of the dressed current operator together with
two adjacent Green functions:
\begin{equation}
\label{J}
 \mathbf{J}_\mathbf{r}
  = F \int \frac{d^2 p}{(2\pi)^2}\, G^A_\mathbf{p} \bm{\sigma} G^R_\mathbf{p} e^{i \mathbf{p} \mathbf{r}}
  = \frac{2 \boldsymbol{\nabla} \bigl[ G^R_\mathbf{r} - G^A_\mathbf{r} \bigr]}{\pi \alpha (\epsilon^2 + 3 m^2)}.
\end{equation}
The last expression holds to the leading order in $\alpha$. The two elements containing $\mathbf{J}$ in each diagram yield $\alpha^{-2}$ that cancels the
contribution of the two crossing impurity lines. Hence upon substitution of Eq.\ (\ref{J}) into Eqs.\ (\ref{XPsi}), all Green functions can be considered in
the clean limit. This also validates the use of the diagonal part $F$ of the dressed current operator (\ref{dressed}) in Eq.~(\ref{J}).

The clean Green function has the following form in real space:
\begin{equation}
 G^{R,A}_\mathbf{r}
  = \frac{1}{4} \bigl[ \epsilon + m \sigma_z - i \bm{\sigma} \boldsymbol{\nabla} \bigr] \bigl[ Y_0(p_0 r) \mp i J_0(p_0 r) \bigr].
\end{equation}
where $J_0$ and $Y_0$ are the Bessel functions of the first and second kind, respectively, and $p_0 = \sqrt{\epsilon^2 - m^2}$ is the Fermi momentum. After
taking the trace and integrating over directions of $\mathbf{r}$, the integrals in Eqs.\ (\ref{XPsi}) reduce to
\begin{subequations}
\label{XPsiint}
\begin{align}
 \sigma_{xy}^\text{X}
  &= \frac{4\pi e^2}{h} \frac{\epsilon m (\epsilon^2 - m^2)}{(\epsilon^2 + 3 m^2)^2} \int ds\, J_1^2 (J_1 Y_0 - J_0 Y_1), \\
 \sigma_{xy}^\Psi
  &= \frac{2\pi e^2}{h} \frac{\epsilon m (\epsilon^2 - m^2)}{(\epsilon^2 + 3 m^2)^2} \int ds\, \bigl[ J_1^2 (J_1 Y_0 + J_0 Y_1) \nonumber \\
  &\qquad\qquad\qquad\qquad\qquad\quad + J_0 J_1 Y_0(J_0-J_2) \bigr],
\end{align}
\end{subequations}
% \begin{subequations}
% \label{XPsiint}
% \begin{align}
%  \sigma_{xy}^\text{X}
%   &= \frac{4\pi e^2}{h} \frac{\epsilon m (\epsilon^2 - m^2)}{(\epsilon^2 + 3 m^2)^2} \int ds\, J_1^2 (J_1 Y_0 - J_0 Y_1), \\
%  \sigma_{xy}^\Psi
%   &= \frac{2\pi e^2}{h} \frac{\epsilon m (\epsilon^2 - m^2)}{(\epsilon^2 + 3 m^2)^2} \int ds\, J_1^2 (J_1 Y_0 + J_0 Y_1),
% \end{align}
% \end{subequations}
where the argument $s = p_0 r$ of the Bessel functions is suppressed. Note that the main contribution to the integrals in Eqs.\ (\ref{XPsiint}) is indeed
provided by $s \sim 1$, i.e., by distances of the order of the Fermi wave length $1/p_0$ as anticipated.

Evaluation of Eqs.~(\ref{XPsiint}) yields the result:
\begin{equation}
 \sigma_{xy}^\text{X}
  = \frac{4 e^2}{h} \frac{\epsilon m (\epsilon^2 - m^2)}{(\epsilon^2 + 3 m^2)^2},
 \qquad
 \sigma_{xy}^\Psi
  = 0.
\end{equation}
We see that the $\Psi$ diagram actually vanishes while the X term gives a positive contribution in the skew-scattering channel. Cancellation of the $\Psi$
diagram is accidental for the massive Dirac model and does not hold for a more general model of AHE. Comparing to the results of Ref.\ \cite{Sinitsyn2006}, we
conclude that the whole skew-scattering part changes sign upon inclusion of the X diagram (see Fig.\ \ref{fig:intsjss}),
\begin{equation}
 \label{ss}
 \sigma_{xy}^\text{ss}
  = \frac{e^2}{h}\, \frac{(\epsilon^2 - m^2) (5\epsilon^2 + 3m^2)}{2\epsilon (\epsilon^2 + 3 m^2)^2}.
\end{equation}
The full result for the anomalous Hall conductivity in the zeroth order in $\alpha$ is given by Eq.~(\ref{DX}).

Thus we have shown that the X diagram contributes to the anomalous Hall effect in the leading order with respect to the disorder strength. (This result has been
anticipated already in Ref.\ \cite{Sinitsyn2007} but the contribution was incorrectly regarded as being parametrically small.) 
While evaluation of the skew scattering on double-impurity requires an accurate quantum-mechanical calculation including X and $\Psi$ diagrams, the
resulting cross-section can be directly used in the collision integral within the quasiclassical Boltzmann kinetic equation. The above result is then reproduced
along the lines of Ref.\ \cite{Sinitsyn2007}.

When three or more intersecting impurity lines are included in the conductivity diagram, they produce an extra smallness leading to the weak localization
correction $\delta \sigma_{xy} \propto \alpha$. The general logic in estimating the order of a diagram is the following. Finite Hall conductivity requires at
least one Green function in the diagram to be taken away from the mass shell $p = p_0$. The impurities connected by this Green function have to be considered as
a single scattering complex. Once these two impurities are combined, counting remaining crossings determines the order of the diagram in the parameter $\alpha$.
Equivalently, the order of the diagram can be estimated from the number of loops in real space after the two close impurities are fused together.

Weak localization correction to the longitudinal conductivity $\sigma_{xx}$ for the model (\ref{model}) was computed in Ref.\ \cite{Kachorovskii}. In that case,
X and $\Psi$ diagrams lead to logarithmic renormalization \cite{Ludwig1994, Aleiner06, Ostrovsky06, Evers08} of $\alpha$ but are excluded from the weak
localization since they do not contribute to magnetoresistance (do not form a loop in real space). Our result for the Hall conductivity Eq.\ (\ref{DX}) does
not explicitly depend on $\alpha$ hence the renormalization of disorder strength is unimportant. Computing weak localization correction to $\sigma_{xy}$,
however, will require inclusion of additional X- and $\Psi$-type diagrams in the Cooperon loop. This will be the subject of a separate publication.

The massive Dirac Hamiltonian (\ref{Ham}) possesses non-trivial topological properties that are manifested in the finite Berry phase $\pi$ accumulated along a
large loop in momentum space. This is the origin of many fascinating topological effects including odd quantization of the Hall conductivity in graphene
\cite{Novoselov2005, Geim2007}, quantum spin-Hall effect in HgTe \cite{Inoue2006}, and topological magnetoelectric response at the surface of topological
insulators \cite{Qi08, Hasan2010rev, Qi2011rev}. It is also intimately related to the fermion doubling. A finite Berry phase requires the presence of the second
valley in the spectrum or a second (opposite) surface in the case of the topological insulator. The Hall conductivity, observable in a transport experiment,
will necessarily include the contribution of both valleys or surfaces \cite{Ostrovsky2007, Koenig14}. Each of these contributions will have the form of
Eq.~(\ref{DX}) possibly with different values of $\epsilon$ and $m$.

Other models of AHE include kinetic (quadratic in momentum) terms in the Hamiltonian. Two most studied models of this type are the ferromagnetic Rashba model
\cite{Nagaosa2010rev} and the Bernevig-Hughes-Zhang model \cite{BHZ}. Bending of the spectrum due to quadratic terms removes the non-trivial total Berry phase
but doubles the Fermi surface. This significantly complicates the calculation of the Hall conductivity \cite{Nunner07} as compared to the massive Dirac model
considered here. Nevertheless, the X and $\Psi$ diagrams must be also included in $\sigma_{xy}$ within these general models. The result of this calculation will
be published elsewhere.

Skew scattering is relevant for a number of related phenomena such as spin-Hall effect \cite{Perel1971}, polar Kerr effect in chiral superconductors
\cite{Xia2006, Xia2008, Kapitulnik2009}, spin-orbit torque \cite{Manchon2009, Garello2013} etc. A proper quantitative description of such effects also requires
inclusion of X and $\Psi$ diagrams.

In conclusion, we have demonstrated that an accurate evaluation of the anomalous Hall conductivity should include the diagrams with two crossed impurity lines,
Fig.\ \ref{fig:bubbles}(b-d), in addition to the standard set of non-crossing diagrams, Fig.\ \ref{fig:bubbles}(a). Despite impurity-line crossing, these diagrams
contribute in the leading order to the skew scattering component of anomalous Hall conductivity. In the model of weak disorder, skew scattering originates
solely from pairs of impurities at a distance smaller or of the order of the electron wavelength. Such double defects require full quantum-mechanical treatment
beyond the commonly employed ladder approximation. We illustrate our findings using AHE with two-dimensional massive Dirac fermions as a model example. Upon
inclusion of X and $\Psi$ diagrams, the anomalous Hall conductivity (\ref{DX}) is considerably suppressed as compared to the non-crossing approximation result
(\ref{DD}). We argue that X and $\Psi$ diagrams are indispensable for all models of AHE as well as for a number of related phenomena.

\acknowledgments
We are grateful to Igor Gornyi, Nikolai Sinitsyn, and Jairo Sinova for helpful discussions. Equations (\ref{RG}) -- (\ref{Z}) were obtained with
support from Russian Science Foundation (Grant No.\ 14-42-00044). The work was supported by the Dutch Science Foundation NWO/FOM 13PR3118 and by the EU Network
FP7-PEOPLE-2013-IRSES Grant No 612624 ``InterNoM''.

\end{document}